\documentclass[journal,10pt]{IEEEtran}
\usepackage{amsmath,amsfonts}
\usepackage{algorithmic}
\usepackage{algorithm}
\usepackage{array}
\usepackage{textcomp}
\usepackage{stfloats}
\usepackage{url}
\usepackage{verbatim}
\usepackage{graphicx}
\usepackage{cite}
\usepackage{setspace}
\usepackage{color}
\usepackage{tabularx}
\usepackage{extarrows}
\usepackage{blkarray}

\usepackage{booktabs}
\usepackage{multirow}
\usepackage{dsfont}
\usepackage{tabularx}
\usepackage[table]{xcolor}
\usepackage{amsthm}
\usepackage{stfloats}
\usepackage{letltxmacro}
\usepackage[caption=false,font=footnotesize,labelfont=rm,textfont=rm]{subfig}
\newcommand{\ra}[1]{\renewcommand{\arraystretch}{#1}}
\newcommand{\figref}[1]{Fig. \ref{#1}}

\usepackage{hyperref}
\hypersetup{hidelinks} %%去红色方框
\hypersetup{
	colorlinks=true,
	linkcolor=blue
}

\begin{document}

%\title{GDM4MIMO: Generative Diffusion Models Enable Massive MIMO Communications}
\title{GDM4MMIMO: Generative Diffusion Models for Massive MIMO Communications}

%\author{Zhenzhou Jin, Huibin Zhou, Yuanshuo Wang, Xinrui Gong, 
%	Xiaofeng Liu, Li You, Xiqi Gao, and Xiang-Gen Xia
%	% <-this % stops a space
%	\thanks{Zhenzhou Jin, Huibin Zhou, Yuanshuo Wang, Xinrui Gong, and Xiqi Gao are with the National Mobile Communications Research Laboratory, Southeast University, Nanjing 210096, China, and also with the Purple Mountain Laboratories, Nanjing 211100, China (e-mail: lyou@seu.edu.cn; zhuyx@seu.edu.cn; xyqiang@seu.edu.cn; wangwj@seu.edu.cn; xqgao@seu.edu.cn).}% <-this % stops a space
%	\thanks{Xiaofeng Liu is with the Huawei Technologies Co., Ltd., Suzhou 518129, China (e-mail: xf_liu@seu.edu.cn ).}
%	\thanks{Xiang-Gen Xia is with the Department of Electrical and Computer Engineering, University of Delaware, Newark, DE 19716 USA (e-mail: xxia@ee.udel.edu).}}
	%\thanks{Christos G. Tsinos is with the National and Kapodistrian University of Athens, Evia, 34400, Greece and also with the University of Luxembourg, Luxembourg City 2721, Luxembourg (e-mail: ctsinos@uoa.gr).}
	%\thanks{Bj\"{o}rn Ottersten is with the University of Luxembourg, Luxembourg City 2721, Luxembourg (e-mail: bjorn.ottersten@uni.lu).}}

\author{Zhenzhou Jin, Li You, Huibin Zhou, Yuanshuo Wang, Xiaofeng Liu, Xinrui Gong, \\ Xiqi Gao, Derrick Wing Kwan~Ng, and Xiang-Gen Xia
	% <-this % stops a space
	\thanks{Zhenzhou Jin, Li You, Huibin Zhou, Yuanshuo Wang, Xinrui Gong, and Xiqi Gao are with the National Mobile Communications Research Laboratory, Southeast University, Nanjing 210096, China, and also with the Purple Mountain Laboratories, Nanjing 211100, China (e-mail: zzjin@seu.edu.cn; lyou@seu.edu.cn; zhouhb@seu.edu.cn; yswang@seu.edu.cn; xinruigong@seu.edu.cn; xqgao@seu.edu.cn).}% <-this % stops a space
	\thanks{Xiaofeng Liu is with the Huawei Technologies Co., Ltd., Suzhou 215100, China (e-mail: lopxiao1@gmail.com).}
	\thanks{Derrick Wing Kwan~Ng is with the School of Electrical Engineering and Telecommunications, University of New South Wales, Sydney, 2052, Australia (e-mail: w.k.ng@unsw.edu.au).}
	\thanks{Xiang-Gen Xia is with the Department of Electrical and Computer Engineering, University of Delaware, Newark, DE 19716 USA (e-mail: xxia@ee.udel.edu).}}

\markboth{}%
{Shell \MakeLowercase{\textit{et al.}}: A Sample Article Using IEEEtran.cls for IEEE Journals}

%Remember, if you use this you must call \IEEEpubidadjcol in the second
%column for its text to clear the IEEEpubid mark.
\maketitle

\begin{abstract}
Massive multiple-input multiple-output (MIMO) offers significant advantages in spectral and energy efficiencies, positioning it as a cornerstone technology of fifth-generation (5G) wireless communication systems and a promising solution for the burgeoning data demands anticipated in sixth-generation (6G) networks. In recent years, with the continuous advancement of artificial intelligence (AI), a multitude of task-oriented generative foundation models (GFMs) have emerged, achieving remarkable performance in various fields such as computer vision (CV), natural language processing (NLP), and autonomous driving. As a pioneering force, these models are driving the paradigm shift in AI towards generative AI (GenAI). Among them, the generative diffusion model (GDM), as one of state-of-the-art families of generative models, demonstrates an exceptional capability to learn implicit prior knowledge and robust generalization capabilities, thereby enhancing its versatility and effectiveness across diverse applications. In this paper, we delve into the potential applications of GDM in massive MIMO communications. Specifically, we first provide an overview of massive MIMO communication, the framework of GFMs, and the working mechanism of GDM. Following this, we discuss recent research advancements in the field and present a case study of near-field channel estimation based on GDM, demonstrating its promising potential for facilitating efficient ultra-dimensional channel statement information (CSI) acquisition in the context of massive MIMO communications. Finally, we highlight several pressing challenges in future mobile communications and identify promising research directions surrounding GDM.

\end{abstract}

\section{Introduction}
\IEEEPARstart{T}{he} development of future sixth-generation (6G) wireless systems has recently sparked significant interest from industry and academia, unveiling immense potential for the future of communication technologies. 6G is poised to deliver heterogeneous services with more stringent requirements, including global coverage, full-utilization of spectrum, support for a wide range of emerging applications, enhanced sensory experiences, comprehensive digital integration, and robust security \cite{6G}. It is expected that 6G will evolve from terrestrial communication to an integrated space-air-ground-sea (SAGS) communication network, ensuring seamless global connectivity and coverage across diverse environments. Moreover, 6G will deeply integrate communications, computing, storage, control, sensing, positioning, artificial intelligence (AI), and big data to cater to various vertical industries. This integration will enable emerging application scenarios such as integrated communication, localization, and sensing, where multi-modal sensory data from multiple sources (e.g., GNSS, BDS, radar, lidar, and cameras) merge to enhance communication quality and faster-varied applications \cite{liu2022survey}. Additionally, 6G will also enrich users' sensory experience through advanced applications like holographic communications and immersive extended reality (XR). One of the critical elements distinguishing 6G is the role of digital twins (DT). In particular, DT will play a crucial role in 6G systems, simulating and optimizing physical systems, which is essential for optimizing network layout and effective resource allocation \cite{jin2024i2i,Digital_Twin}. As we explore these future possibilities, it is also important to examine current advancements, which will pave the way for 6G's realization, such as the development of massive multiple-input multiple-output (MIMO) technologies.

Massive MIMO serves as a fundamental enabling technology in fifth-generation (5G) systems and is expected to assume an even more pivotal role in the development of 6G networks, particularly with the rise of extremely large-scale MIMO (XL-MIMO). For instance, the prototype Giga-MIMO system from Qualcomm features 4,096 antenna elements and 256 digital links, representing recent significant advancements in massive MIMO\footnote[1]{https://www.rcrwireless.com/20240325/6g/giga-mimo-is-the-foundation-for-wide-area-6g}. In addition, the Blocker-Tolerant millimeter (mm)-Wave MIMO receiver for 5G and beyond, developed by MIT, achieves an error vector magnitude (EVM) of -32.6 dB under the desired signal conditions of receiving a wideband 256-QAM 100 Mega-symbols-per-second (MS/s) signal \cite{MIT01}. In practice, massive MIMO system equips large-scale antenna arrays at the base station (BS) to capitalize on the spatial dimension of wireless resources, substantially boosting both the spectral and energy efficiencies of communication systems significantly. Indeed, the successful implementation of massive MIMO relies heavily on the technical support provided by beamforming, antenna diversity, and spatial multiplexing. Specifically, beamforming optimizes the signal's directivity and strength, ensuring the formation of a high-gain beam towards the target direction. Meanwhile, antenna diversity and spatial multiplexing enable the receiver to select the best signal to improve signal quality and allow independent data streams to be transmitted and received on the same frequency resources to increase information transmission rates, respectively. In the context of massive MIMO technology, it is imperative to address several critical issues. The first issue involves how to reduce the pilot signaling overhead in channel estimation as the channel dimension increases substantially. In addition, it is essential to reduce the implementation complexity, particularly concerning signal processing. Furthermore, non-ideal factors should also be considered to ensure optimal performance and reliability. It is noteworthy that with the continuous advancements in AI, a growing number of deep learning networks and models, particularly task-oriented generative AI (GenAI) with robust implicit prior learning capabilities, are being extensively explored to tackle these pressing challenges.

\begin{figure}[!t]
	\centering
	\includegraphics[width=0.49\textwidth]{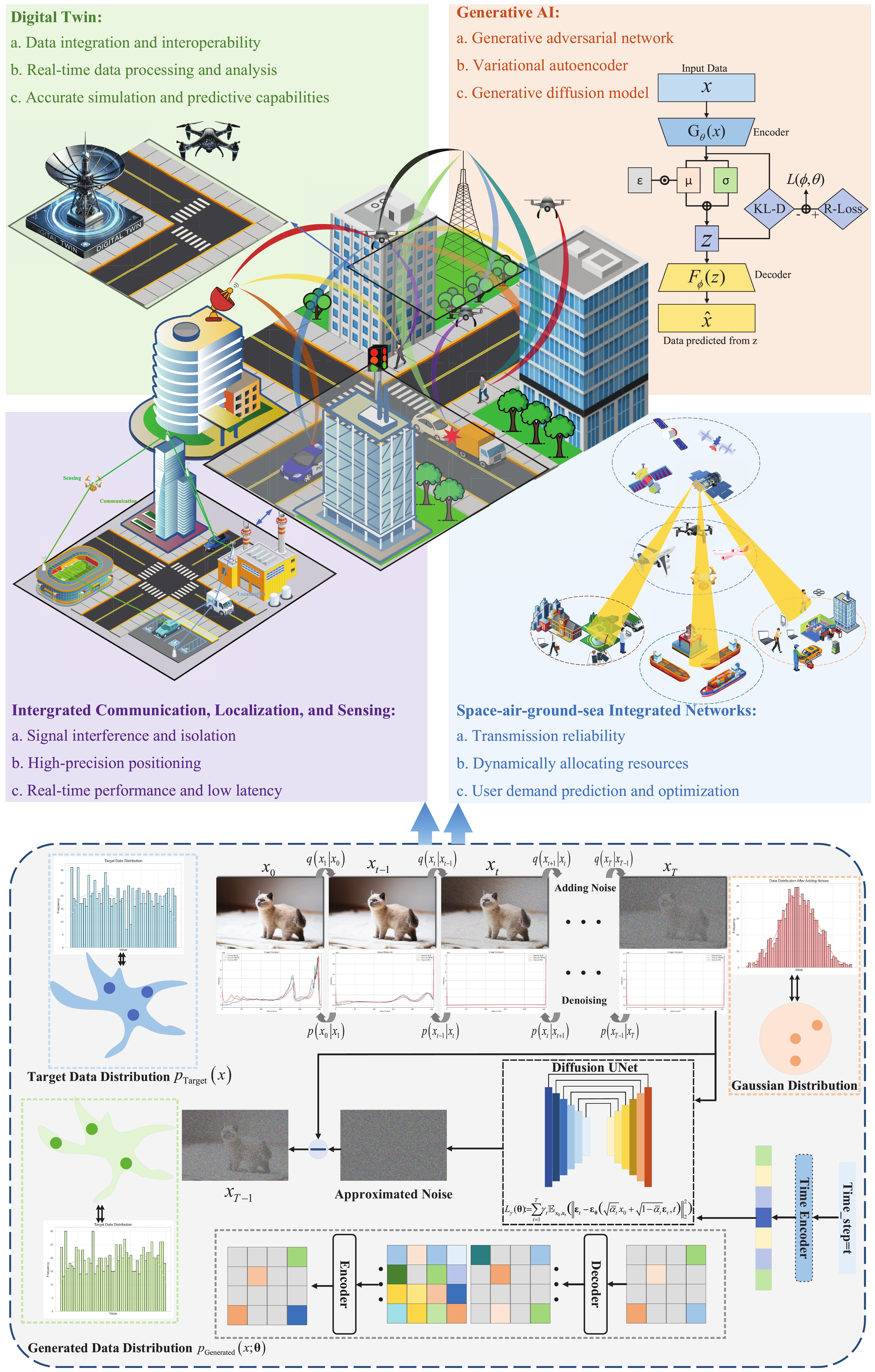}
	\captionsetup{font=footnotesize}
	%\vspace{-0.25cm}
	\caption{The working principles of GDM and its potential to enable several cutting-edge application scenarios in massive MIMO communications, including space-air-ground-sea integrated networks, integrated communication, localization, sensing, and digital twin, are showcased.}
	%\vspace{-0.2cm}
	\label{fig: frame}
\end{figure}

Deep generative foundation models (GFMs) have unveiled another significant field in AI, propelling us into the era of all-encompassing Artificial Intelligence for General Creativity (AIGC) by effectively capturing and learning the distribution of massive amounts of task-oriented data. Among these, generative diffusion model (GDM) represents a state-of-the-art family of deep generative models, achieving remarkable results in critical fields such as computer vision (CV) and natural language processing (NLP). Recently, various prominent task-oriented models have emerged, including DALL·E 3 and GPT by OpenAI, Imagen by Google Brain, Stable Diffusion by Heidelberg University, and others \cite{Rombach_2022_CVPR}. Specifically, the GDM architecture consists of two interrelated processes: a predefined forward process that transforms the target data distribution into a simpler prior distribution, typically Gaussian, by introducing a fixed amount of noise; and a corresponding reverse process that utilizes a trained neural network to gradually reverse the influence of the forward process, known as ``denoising''. This denoising process often involves simulating ordinary or stochastic differential equations (ODE/SDE) \cite{song2020denoising}. To a certain extent, GDM learns the implicit prior distribution of the target data and eliminates out-of-distribution (OOD) components through the aforementioned processes. This enables GDM to achieve robust ``denoising'' and ``generation'' capabilities. Upon deeper reflection, these inherent properties of ``denoising'' and ``generation'' within GDMs align closely with a promising objective in massive MIMO communications: effectively filtering noise and reconstructing informative signals from noisy ones. Detailed insights into the working mechanism of GDM and its potential to enable several cutting-edge application scenarios in the realm of massive MIMO communications are showcased in \figref{fig: frame}.

All the aforementioned aspects have motivated us to pursue this study. Importantly, the integration of GDM into the context of massive MIMO communications remains nascent, and our objective in this paper is to elucidate the key characteristics of GDMs and delve into potential research directions for their applications within the field of massive MIMO communications. The rest of the paper is organized as follows. In Section II, an overview of the background of massive MIMO, the framework of GFMs, and the fundamentals of GDM are provided. In Section III, the recent research advances, including non-ideal massive MIMO wireless communications and massive MIMO CSI acquisition, are presented. In Section IV, several open issues and promising research directions related to GDM for massive MIMO communication are discussed. Finally, the conclusion is presented in Section V.

\section{Overview of Massive MIMO, GenAI, and GDM}
In this section, we primarily elucidate the fundamentals and concepts behind massive MIMO, GenAI, and GDM, which form the foundation for exploring how GDM, as one of the GenAI models, can be effectively applied to massive MIMO.
\subsection{Background of Massive MIMO}
Massive MIMO stands as one of the most pivotal advancements in the enhancement of spectral efficiency in 5G wireless communications. Since its initial conceptualization by Marzetta, this technology has garnered widespread adoption and implementation in the field. The advantages of massive MIMO technology in communication encompass not only immense spectral efficiency and communication reliability, but also high energy efficiency and favorable propagation characteristics. Additionally, it significantly enhances network capacity and coverage, reduces interference, improves user experience, and supports a greater number of user and device connections, thus addressing the ever-increasing future communication demands.

In recent years, researchers have conducted extensive studies to address the unique challenges in massive MIMO. Firstly, the traditional orthogonal pilot method leads to a linear increase in pilot signaling overhead in high-dimensional channels. As such, some studies utilize the sparsity of the angular domain, sparse clustering Bayesian learning methods, as well as compressed sensing (CS) approaches such as orthogonal matching pursuit (OMP) and other sparsity techniques to effectively achieve channel estimation with low pilot overhead \cite{9693928}. Secondly, a reasonable precoder is designed for massive MIMO downlink (DL) transmission, effectively mitigating inter-user interference. Yet, precoding generally involves matrix multiplication and matrix pseudo-inversion, resulting in significant computational complexity. For example, the computational complexity of the weighted minimum mean square error (WMMSE) algorithm is cubic in the number of BS antennas, which becomes unacceptable in massive MIMO systems. Some studies investigate reducing computational complexity by transforming constrained problems in Euclidean space to unconstrained ones on manifolds, avoiding large matrix inversions through the use of Riemannian optimization. Subsequently, the deployment of practical communication systems cannot always guarantee the ideal working condition, leading to non-ideal hardware impairment (HI) such as low-precision quantization in power amplifiers and nonlinear phenomena. Some studies have applied the Bussgang theory to establish a linear quantization model for digital-to-analog converter/analog-to-digital converter to alleviate the adverse effects of low-precision quantization. Some researchers also design pragmatic schemes to maximize energy efficiency by considering power-consumption outage (PCO) and total power constraints to solve the heat dissipation problem. Meanwhile, by characterizing the statistical properties of nonlinear distortion, the realization and rate lower bound of systems with HI can be derived.

\subsection{Framework of GFMs}
In contrast to conventional artificial intelligence models, commonly referred to as discriminative AI, which predominantly focus on analyzing or classifying existing data, GenAI can effectively learn to model and generate target data by learning the intricate structure and characteristics of the data through an implicit prior distribution. This encompasses the generation of fresh textual content, auditory material, visual imagery, and even synthetically generated time-series data. GenAI is not confined to specific tasks or domains; rather, it can be broadly applied to a wide range of intricate creative and problem-solving processes. The proficiency of GenAI in analyzing structured data distributions makes it a promising tool for addressing challenges in massive MIMO systems. For instance, GenAI can function as an equilibrium estimator, acquiring knowledge of the mapping from corrupted signals to transmitted signals, and inferring the clean signal from the corrupted observations at the receiver.

Recently, various GenAI models have been applied to NLP and CV, demonstrating impressive performance. For example, generative adversarial networks (GANs) generate simulation data through a competition between generators and discriminators to compete against each other. On the other variational autoencoder (VAE) combines the benefits of autoencoders and probabilistic graphical models to generate new data samples by learning the latent representations of the data. In addition, flow-based generative model (FGM) converts a complex probability distribution into a simplified one, subsequently mapping it back to the original data space through a series of reversible transformations. Furthermore, the generative transformer model (GTM) can capture long-distance dependency relationships, learn the mapping relationship between input and output sequences, and generate the target output sequence. Also, GPT-4 is a representative application of GTM to large language model (LLM)\cite{achiam2023gpt}. GDM can fit complex target distribution using a fixed forward process and learnable reverse process. A comparison summary of different GenAI models is shown in Table \ref{tab:t1}.

\newcolumntype{L}{>{\hspace*{-\tabcolsep}}l}
\newcolumntype{R}{c<{\hspace*{-\tabcolsep}}}
\definecolor{lightblue}{rgb}{0.93,0.95,1.0}
\begin{table}[htbp]
	\captionsetup{font=footnotesize}
	%\vspace{-0.2cm}
	\caption{Comparison of different GFMs}
	\centering
	\setlength{\tabcolsep}{2.6mm}%宽度
	\ra{1.3}%高度
	\label{tab:t1}    %表格标签
	\scriptsize
	%\rowcolors{1}{lightblue}{white}
	\scalebox{0.92}{\begin{tabular}{LccR}
			%\scriptsize
			\toprule
			Model &  Architecture &  Advantage & Disadvantage\\
			\midrule
			%Earth radius $R_e$ && 6378 $(\text{km})$\\
			\rowcolor{lightblue}
			& Two neural networks, & High-quality generated & \\
			\rowcolor{lightblue}
			\multirow{-2}*{GAN} & generator and discriminator &  data  & \multirow{-2}*{Mode collapse} \\

			&   &  High-diversity of  &   \\
			& Encoder-decoder structure &  generated data and  &   \\
			& with variational inference &   coherent sample  &   \\
			\multirow{-4}*{VAE}&  &   generation & \multirow{-4}*{Low sample quality} \\
			
			%\rowcolor{lightblue}
			%&  &  & High training cost  \\
			%\rowcolor{lightblue}		
			%&  \multirow{-2}*{Coupling layer with}   & %\multirow{-2}*{Reliable estimation of}  &  and difficult to \\	
			%\rowcolor{lightblue}
			%\multirow{-3}*{FGM}&\multirow{-2}*{invertible transformation}&  \multirow{-2}*{the data density}& model\\	
			
			\rowcolor{lightblue}
			& Coupling layer with & Reliable estimation of & High training cost  \\
			\rowcolor{lightblue}
			\multirow{-2}*{FGM}  &  invertible transformation & the data density & and low adaptability\\

			&   & Robust reasoning and & Numerous sampling  \\	
			& \multirow{-2}*{Fixed forward process and} & effective prior  &  steps and extended \\		
			\multirow{-3}*{GDM}& \multirow{-2}*{learnable reverse process} &  information learning & sampling duration \\							
			\bottomrule
		\end{tabular}
	}
\end{table}

\subsection{Fundamentals of GDM}
The strong theoretical foundation and explainability of GDM bring new ideas and methods to the field of GenAI. As a matter of fact, GDM has shown strong performance in various tasks such as image generation and text generation. The popularity of stable diffusion is a specific implementation of GDM in the image field, which can generate images based on various text prompts\cite{Rombach_2022_CVPR}. GDM combines concepts from stochastic processes, Markov chains, and probability theory, enhancing its explainability and theoretical grounding. The core idea behind GDM is to fit a complex target distribution from a simple distribution (usually Gaussian noise) through a series of reverse diffusion steps. By optimizing the parameters of the diffusion model, the desired samples are gradually generated. GDM can be divided into two main stages: the forward process and the reverse process.

The forward process in GDM resembles Brownian motion with a time-varying coefficient. It involves introducing subtle noise to the input data at each iteration, with the noise scale varying dynamically at every step. As the training data undergoes a gradual deterioration, it ultimately transitions into pure Gaussian noise. Fitting the probability model of the reverse process proves challenging, as it necessitates learning the distribution of all potential data samples to compute the reverse conditional probability accurately. Usually, the reverse process employs a trained neural network. The primary objective of the training process is to minimize the difference between the fitted complex distribution and the target distribution, typically achieved by minimizing the variational boundary on negative log-likelihood exploiting Kullback-Leibler (KL)-divergence. It should be pointed out that compared with VAE and GAN, the trained GDM uses the learned prior distribution to gradually restore the complex target distribution in the reverse process, ensuring high quality and strong robustness of samples. During the reverse process, extensive performance evaluations are required, resulting in slower sampling speeds. Recent studies have brought forth advancements such as distillations and pruning to enhance the sampling rate and overall performance of the models.
%\subsection{Massive MIMO Communication Systems}
%\subsection{Advent of the GenAI Era}
%\subsection{Fundamentals of GDM}
%\subsection{Differences between GDM and other generative models}

\section{Recent Research Advances}
In this section, we delve into several potential application scenarios within massive MIMO wireless communication systems and present some recent research advancements that leverage the powerful learning capabilities of GDM for harnessing the implicit prior information of target data.

%\subsection{Eliminating Channel Noise Resulting from Non-Ideal Factors}
\subsection{Non-ideal Massive MIMO Wireless Communications}
The spectral efficiency of massive MIMO systems is inherently constrained by the information-theoretic capacity, which primarily relies on the limitations imposed by non-ideal transmission factors. In practical implementations of massive MIMO communication, deviations from theoretical models are common. These include channel modeling errors, channel statement information (CSI) acquisition errors, limitations in computational resources, interference from unintended communication links, and environmental noise. In addition, future massive MIMO systems aim to prioritize cost-effective and eco-friendliness deployment strategies. However, this pursuit will easily lead to impairments in the transceiver, such as amplifier nonlinearities, quantization errors, phase noise, and I/Q imbalance. These non-ideal factors can significantly degrade the spectral efficiency of wireless links, thereby affecting the overall efficacy of wireless communications.

Notably, the errors caused by actual non-ideal factors can be viewed as noise. By leveraging the powerful implicit prior learning ability of GDM to discern the distribution of target data, and then detecting and eliminating OOD samples (a process known as denoising), this approach can effectively alleviate these issues. The study by \cite{letafati2023denoising} initially explored the deployment of GDM within wireless communication frameworks, considering practical scenarios like hardware impairments (HI), low SNR, and quantization errors. Instead of designing a communication system to avoid HI and estimation/decoding errors, they train a GDM to counteract these distortions, thereby introducing ``inherent resilience'' into massive MIMO communication systems. More recently, the authors of \cite{10480348} proposed a GDM-based channel denoising model for semantic communication over the wireless channel. After channel equalization, this model can be embedded into the physical layer as a new module, capable of learning the actual distribution of channel input signals, and subsequently utilizing the acquired prior knowledge to eliminate channel noise, promising to enhance communication performance.
\subsection{Massive MIMO CSI Acquisition}
The efficacy of numerous techniques in massive MIMO systems, such as beamforming and resource allocation, critically depends on the accurate acquisition of CSI. Nevertheless, in massive MIMO orthogonal frequency division multiplexing (OFDM) systems, CSI acquisition faces several tricky challenges. Specifically, the rapid increase in the dimension of the base station antenna array, the adoption of hybrid digital-analog architecture, the increase in the number of user terminals, and other practical factors result in a prohibitively high pilot signaling overhead required for accurate CSI acquisition. Therefore, it is paramount to achieve accurate CSI acquisition while minimizing pilot signaling overhead and complexity costs. Remarkably, CSI acquisition aligns well with the powerful implicit prior learning capability of GDM, offering a novel methodology for CSI acquisition.

Recent studies have explored the implicit prior learning capability of GDM, especially investigating its application in acquiring CSI in massive MIMO systems. The authors of \cite{fesl2024diffusion} employed a top-rated GDM as a prior information generator, continuously learning the MIMO channel distribution in the sparse angular domain during training. Then, by employing an sampling strategy that scales down reverse diffusion steps for lower SNR than the given pilot observations, resulting in a GDM-based estimator that achieves low complexity overhead. In addition, the authors of \cite{arvinte2022mimo} proposed a GDM based on posterior sampling, which utilizes a denoising score-matching framework to learn the gradient of the log-prior distribution of MIMO channels during training, referred to as the score. Then, during the inference process, the trained GDM and the received pilot signals are employed to iteratively update the MIMO channel through multiple rounds of posterior sampling. These aforementioned GDM-based CSI acquisition approaches demonstrate substantial potential to enhance estimation accuracy and generalization capabilities. Therefore, such studies indicate that GDM holds promising potential for facilitating CSI acquisition in the context of massive MIMO communication.

\begin{figure}[!t]
	\centering
	\includegraphics[width=0.46\textwidth]{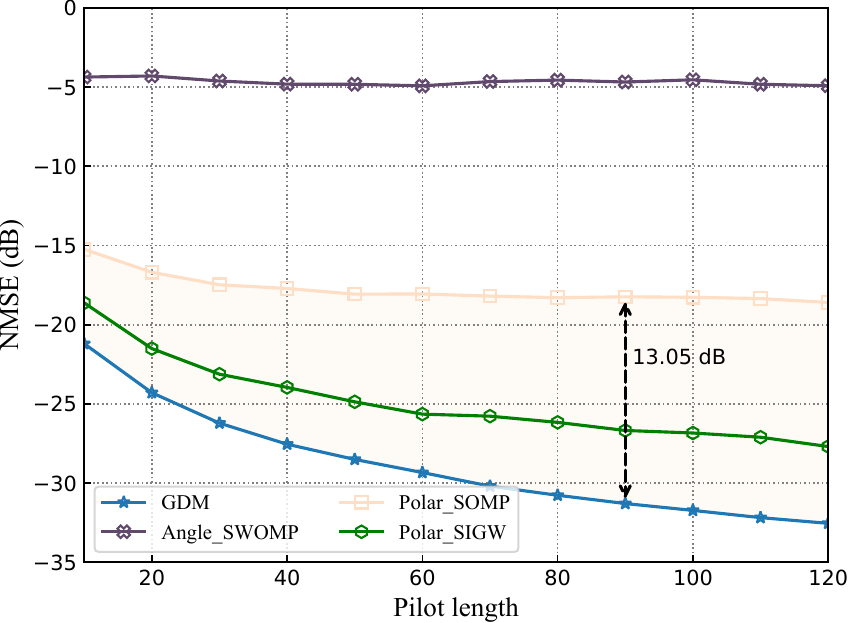}
	\captionsetup{font=footnotesize}
	%\vspace{-0.25cm}
	\caption{NMSE performance of CSI acquisition utilizing GDM.}
	%\vspace{-0.2cm}
	\label{fig:nmse}
\end{figure}

%\subsection{Precoding}
%\subsection{Multi-Satellite-Enabled ISAC}
%\section{(\textit{JZZ}) CASE STUDIES}
\textit{Case Study:} Given the pivotal role of precise CSI in bolstering the communication efficacy of massive MIMO systems, and the advent of XL-MIMO is identified as an essential enabler for future 6G networks. In response to this, our study further ventures into GDM-based XL-MIMO near-field CSI acquisition, aiming to cater to the burgeoning data rate requisites anticipated in 6G landscapes. Specifically, we consider a multi-user uplink scenario within an XL-MIMO OFDM system. In this configuration, the BS is equipped with a uniform linear array (ULA) comprising $N=256$ antenna elements, which are arranged with half-wavelength spacing. The system serves $M=16$ users, each utilizing a single-antenna device. To manage the system complexity and hardware constraints, the BS employs a hybrid digital-analog architecture, incorporating a limited number of RF chains, denoted as $N_{\rm{RF}}=16$. In our analysis, we employ three classical CSI acquisition algorithms, including angle-domain simultaneous weighted orthogonal matching pursuit (SWOMP), polar-domain simultaneous orthogonal matching pursuit (SOMP), and simultaneous iterative gridless weighted (SIGW) \cite{9693928}, as benchmarks and contrast them with the GDM-based approach. \figref{fig:nmse} showcases the normalized mean squared error (NMSE) performance of CSI acquisition versus pilot overhead based on different approaches. This case study demonstrates that the GDM-based channel estimation significantly outperforms the compared algorithms and exhibits robust generalization capabilities, which is anticipated to enhance the efficiency of CSI acquisition. 

%\begin{figure}[!t]
%	\centering
%	\subfloat[User terminals are located at (5 m, 35 m).]{\includegraphics[width=0.38\textwidth]{figures/magazine_R(5,35)_pilots_snr_r=10_8db_7.eps}\label{fig:near}}
%	\hfill
%	\subfloat[User terminals are located at (350 m, 400 m).]{\includegraphics[width=0.38\textwidth]{figures/magazine_R(350,400)_pilots_snr_r=10_8db_7.eps}\label{fig:far}}
%	\caption{NMSE performance of CSI acquisition in relation to pilot overhead.
%		(a) User terminals are located at (5 m, 35 m), which belongs to the near-field communication area within the considered system. (b) User terminals are located at (350 m, 400 m), which belongs to the far-field communication area within the considered system.}
%	\label{fig:nmse}
%\end{figure}

%\subsection{XL-MIMO Near-Field CE}

%\subsection{RIS-Aided Massive MIMO CE}

%\section{Challenges and Key Technologies}
%\subsection{Challenges}
%\subsection{Key Technologies}

\section{Open Issues and Research Opportunities}
The previous sections provided a comprehensive overview of GDM, covering its definition, enabling technologies, frameworks, and applications and concluding with a case study, particularly highlighting its application in massive MIMO systems. New scenarios and applications will inevitably emerge with the introduction of SAGS communication, DT, and integrated communication, localization, and sensing in 6G wireless systems. However, these advancements will also introduce a set of significant challenges that must be addressed. We believe that the following potential directions are particularly noteworthy.

\subsection{Space-Air-Ground-Sea Integrated Network}
The integrated communication system for SAGS environments aims to achieve seamless connectivity and information exchange across these environments. However, implementing such an integrated communication system faces numerous challenges, particularly in maritime and aerial environments. The primary difficulty lies in the need for ground stations; both maritime and aerial regions need ground-based infrastructure, rendering traditional ground-based communication networks ineffective. Consequently, relying solely on ground-based communication methods is impractical for these environments. To overcome this challenge, satellite communication has emerged as a pivotal technology for achieving integrated communication across maritime, terrestrial, aerial, and space environments. Despite its critical role, satellite communication also encounters issues such as latency, bandwidth limitations, and adaptability to dynamic environments.

As an advanced signal processing technology, GDM is applied in satellite communication to address various challenges that traditional communication technologies cannot overcome. Implementing GDM in satellite communication primarily involves signal enhancement, prediction, and data reconstruction. Specifically, GDM trains on extensive historical communication data to learn various interference and attenuation patterns during signal transmission. This training enables it to predict and enhance current signal transmission conditions, improving signal strength and cleanliness. Additionally, GDM leverages its generative capabilities to reconstruct signals disrupted or partially lost during transmission, ensuring data integrity and reliability. This method is crucial for long-distance transmission and communication in harsh environments, significantly enhancing communication quality and stability.

\subsection{Integrated Communication, Localization, and Sensing}
Integrated communication, localization, and sensing combine wireless communication, environmental sensing, and precise positioning functions into a unified system that utilizes shared wireless resources and infrastructure to achieve more efficient and intelligent operations \cite{loc}. In such a system, wireless signals are exploited for both data transmission and detecting and measuring environmental information, thereby providing comprehensive positioning and sensing services. Signal interference and isolation are critical challenges in these systems. Since communication and sensing signals share the same spectrum resources, mutual interference can occur, resulting in compromised system performance. 

To address these challenges, GDM plays a crucial role in enhancing the performance of integrated systems. Diffusion models can generate both realistic clean signals and signals with interference for system training. By leveraging these generated samples, communication systems can undergo adversarial training to enhance their robustness and anti-interference capabilities in challenging environments. Additionally, the signal data generated by GDM can be utilized to develop adaptive interference cancellation algorithms, enabling the system to dynamically adjust to different interference scenarios, thereby improving the overall performance and reliability of the integrated communication, localization, and sensing system. This approach not only enhances the precision and efficiency of signal processing but also lays the foundation for achieving a more intelligent and efficient integrated system.

\subsection{Digital Twin}
The seamless integration of the digital and physical realms is the cornerstone of the 6G vision. As 6G networks emerge, accommodating a myriad of connected devices and facilitating large-scale Internet-of-Things (IoT) deployments, the need for tools to create accurate digital representations of the vast and complex real-world scenarios 6G networks inhabit becomes more pronounced \cite{Digital_Twin2}. However, the transformation of vast, diverse, dynamic sensory data into precise, actionable DT models poses significant challenges. These challenges include handling high data volumes, ensuring real-time data processing, and maintaining the accuracy and fidelity of the digital twins. 

GDM emerges as a potent solution to these challenges effectively, intertwining the promise of 6G with the efficacy of DT. Specifically, GDM can generate synthetic data that mirrors real-world conditions, filling gaps where actual sensory data might be sparse or missing. This capability ensures that digital twin models are robust and comprehensive. With their ability to efficiently process large-scale datasets, GDM supports real-time updates to DT, reflecting the current state of the physical environment.
Furthermore, GDM can predict future states of the network and environment based on current and historical data, allowing for proactive management and optimization of 6G networks. They also excel at detecting anomalies or irregularities by learning the typical patterns of network behavior and environmental conditions, thus mitigating potential issues before they escalate. By continuously learning and adapting from new data, GDM improves their accuracy and relevance, ensuring that DT remains precise representations of their physical counterparts. By leveraging these capabilities, GDM transforms the vast amounts of sensory data collected by 6G networks into precise, actionable DT models. This enhances the functionality and efficiency of 6G networks and paves the way for innovative applications and services that rely on accurate digital representations of the physical world. Thus, GDM serves as a bridge, enabling the seamless integration of the digital and physical realms in the context of 6G and beyond.
%\subsection{Space-Air-Ground-Sea Integrated Network}
%\subsection{Massive MIMO ISAC}
%\subsection{Digital Twin}
%\subsection{Next Generation MIMO}
%\subsection{Holographic Metasurface \& Fluid Antenna assisted Massive MIMO CE} 

%\subsection{Novel Antenna Architecture Design}
%\subsection{Space-Air-Ground-Sea Integrated Network (SAGSIN) Security With ISAC}
%\subsection{Novel Waveform Design}

\section{Conclusion}
In this paper, we explored generative foundation models and uncovered its application potential in massive MIMO wireless communications. Specifically, we first introduced the massive MIMO communication system, GenAI, and GDM. We then analyzed their robust implicit prior learning capabilities and generalization abilities through GDM's distinctive ``noise addition'' and ``noise removal'' processes. Additionally, two primary challenges in massive MIMO communications were identified: non-ideal massive MIMO wireless communications and massive MIMO CSI acquisition, with solutions based on GDM presented. Note that we emphasized the limitless potential of GDM in massive MIMO CSI acquisition through the showcased study case. Finally, we present several unresolved issues and research directions related to the space-air-ground-sea integrated network, as well as integrated communication, localization, sensing, and digital twins, all of which can be explored through GDM.

\bibliographystyle{IEEEtran}  
\bibliography{reference}

\end{document}